\pdfoutput=1
\documentclass[journal=jacsat,manuscript=article]{achemso}
\usepackage{chemformula} 
\usepackage[T1]{fontenc} 
\usepackage{graphicx}
\usepackage{caption}
\usepackage{subcaption}
\usepackage{pdfpages}

\author{Iman Esmaeil Zadeh\footnote[0]{$\ast$ These authors contributed to this work equally}}
\affiliation[TU delft]{Optics Research Group, ImPhys Department, Faculty of Applied Sciences, Delft University of Technology, Delft 2628 CJ, The Netherlands.}
\alsoaffiliation[SQ]
{Single Quantum B.V., Delft 2628 CJ, The Netherlands.}
\email{i.esmaeilzadeh@tudelft.nl}

\author{Johannes W. N. Los $^\ast$}
\affiliation[SQ]
{Single Quantum B.V., Delft 2628 CJ, The Netherlands.}

\author{Ronan B. M. Gourgues}
\affiliation[SQ]
{Single Quantum B.V., Delft 2628 CJ, The Netherlands.}

\author{Jin Chang}
\affiliation[TU delft]{Optics Research Group, Imphys Department, Faculty of Applied Sciences, Delft University of Technology, Delft 2628 CJ, The Netherlands.}

\author{Ali W. Elshaari}
\affiliation[KTH]
{Quantum Nano Photonics Group, Department of Applied Physics, Royal Institute of Technology (KTH), Stockholm 106 91, Sweden}

\author{Julien Zichi}
\affiliation[KTH]
{Quantum Nano Photonics Group, Department of Applied Physics, Royal Institute of Technology (KTH), Stockholm 106 91, Sweden}

\author{Yuri J. van Staaden}
\affiliation[TU delft]{Optics Research Group, Imphys Department, Faculty of Applied Sciences, Delft University of Technology, Delft 2628 CJ, The Netherlands.}

\author{Jeroen Swens}
\affiliation[TU delft]{Optics Research Group, Imphys Department, Faculty of Applied Sciences, Delft University of Technology, Delft 2628 CJ, The Netherlands.}

\author{Nima Kalhor}
\affiliation[SQ]
{Single Quantum B.V., Delft 2628 CJ, The Netherlands.}

\author{Antonio Guardiani}
\affiliation[SQ]
{Single Quantum B.V., Delft 2628 CJ, The Netherlands.}

\author{Yun Meng}
\affiliation[Tianjin-uni]{School of Precision Instrument and Optoelectronic Engineering, Tianjin University, Tianjin 300072, China.}
\alsoaffiliation[Tianjin-uni2]
{Key Laboratory of Optoelectronic Information Science and Technology, Ministry of Education, Tianjin 300072, China.}

\author{Kai Zou}
\affiliation[Tianjin-uni]{School of Precision Instrument and Optoelectronic Engineering, Tianjin University, Tianjin 300072, China.}
\alsoaffiliation[Tianjin-uni2]
{Key Laboratory of Optoelectronic Information Science and Technology, Ministry of Education, Tianjin 300072, China.}

\author{Sergiy Dobrovolskiy}
\affiliation[SQ]
{Single Quantum B.V., Delft 2628 CJ, The Netherlands.}

\author{Andreas W. Fognini}
\affiliation[SQ]
{Single Quantum B.V., Delft 2628 CJ, The Netherlands.}

\author{Dennis R. Schaart}
\affiliation[TUDRI]{Medical Physics \& Technology, Radiation Science \& Technology department, Faculty of Applied Sciences, Delft University of Technology}

\author{Dan Dalacu}
\affiliation[NRC]{National Research Council of Canada, Ottawa, ON K1A 0R6, Canada.}
\author{Philip J. Poole}
\affiliation[NRC]{National Research Council of Canada, Ottawa, ON K1A 0R6, Canada.}

\author{Michael E. Reimer}
\affiliation[IQC]{Institute for Quantum Computing and Department of Electrical \& Computer Engineering, University of Waterloo, Waterloo, ON N2L 3G1, Canada.}

\author{Xiaolong Hu}
\affiliation[Tianjin-uni]{School of Precision Instrument and Optoelectronic Engineering, Tianjin University, Tianjin 300072, China.}
\alsoaffiliation[Tianjin-uni2]
{Key Laboratory of Optoelectronic Information Science and Technology, Ministry of Education, Tianjin 300072, China.}

\author{Silvania F. Pereira}
\affiliation[TU delft]{Optics Research Group, Imphys Department, Faculty of Applied Sciences, Delft University of Technology, Delft 2628 CJ, The Netherlands.}

\author{Val Zwiller}
\affiliation[KTH]
{Quantum Nano Photonics Group, Department of Applied Physics, Royal Institute of Technology (KTH), Stockholm 106 91, Sweden}
\alsoaffiliation[SQ]
{Single Quantum B.V., Delft 2628 CJ, The Netherlands.}

\author{Sander N. Dorenbos}
\affiliation[SQ]
{Single Quantum B.V., Delft 2628 CJ, The Netherlands.}

\title[An \textsf{achemso} demo]
  {A platform for high performance photon correlation measurements}


\begin{document}

\renewcommand{\floatpagefraction}{.8}%

\begin{abstract}

 A broad range of scientific and industrial disciplines require precise optical measurements at very low light levels. Single-photon detectors combining high efficiency and high time resolution are pivotal in such experiments. By using relatively thick films of NbTiN (8-11\,nm) and improving the pattern fidelity of the nano-structure of the superconducting nanowire single-photon detectors (SNSPD), we fabricated devices demonstrating superior performance over all previously reported detectors in the combination of efficiency and time resolution. Our findings prove that small variations in the nanowire width, in the order of a few nanometers, can lead to a significant penalty on their temporal response. Addressing these issues, we consistently achieved high time resolution (best device 7.7\,ps, other devices $\sim$10-16\,ps) simultaneously with high system detection efficiencies ($80-90\%$) in the wavelength range of 780-1000\,nm, as well as in the telecom bands (1310-1550\,nm). The use of thicker films allowed us to fabricate large-area multi-pixel devices with homogeneous pixel performance. We first fabricated and characterized a $100\times100\, \mu m^2$ 16-pixel detector and showed there was little variation among individual pixels. Additionally, to showcase the power of our platform, we fabricated and characterized 4-pixel multimode fiber-coupled detectors and carried out photon correlation experiments on a nanowire quantum dot resulting in $g^2(0)$ values lower than 0.04. The multi-pixel detectors alleviate the need for beamsplitters and can be used for higher order correlations with promising prospects not only in the field of quantum optics, but also in bio-imaging applications, such as fluorescence microscopy and positron emission tomography.  

\end{abstract}

\section{Introduction}

SNSPDs have already pushed the limits in several fields such as CMOS testing \cite{Zhang:2003}, biomedical imaging \cite{Peer:2007}, laser ranging \cite{McCarthy:2013}, and quantum communication \cite{Yin:2013,Vallone:2016}. These detectors have unparalleled performance: high efficiency ($>$90\%) \cite{Marsili:2013, Zhang:2016, Zadeh:2017}, time resolution ($<$15\,ps)\cite{Zadeh:2017,Wu:2017}, and count-rate \cite{Huang:2018}. Yet, for studying fast phenomena in chemistry, biology, physics, detectors with better timing jitter are required. Although direct high time-resolution measurements are possible, for example with streak cameras, they suffer from low detection efficiencies typically < 10\% and high dark count rates \cite{Muler2007,StreakCamera}, moreover they are bulky and costly. Meanwhile, the fiber coupled SNSPDs allow for a cost efficient integration of many high performance detectors in a single cryostat. Here we demonstrate SNSPDs made of relatively thick NbTiN films with superior time resolution and high efficiencies. Moreover, we show large area sensors using arrays of SNSPDs with minimal variation in performance over the array. In addition to optical imaging applications, this technology could be used to read-out fast scintillators, potentially enabling the detection of ionizing radiation with unprecedented time resolution, which is of great interest for time-of-flight positron emission tomography \cite{Seifert:2012}. We showcase our platform by carrying out photon correlation measurements using a multi-pixel multimode fiber-coupled detector. 

We fabricate our detectors from sputtered NbTiN films on top of a $\mathrm{SiO_2/Au}$  cavity or a distributed Bragg reflector. To achieve the highest possible absorption in the superconducting layer, increasing the critical current and reducing kinetic inductance (and hence the deadtime), thicker films are desirable \cite{Zadeh:2017}. On the other hand, increasing the film thickness leads to a higher energy gap and reduction of the kinetic energy of the quasiparticles. Therefore, for such films, it is challenging to reach saturation of the internal efficiency, and hence, both the material properties of the film and the fabrication uniformity of the SNSPD nanowire meander have to be controlled.

\section{Fabrication and Characterization}

\begin{figure}[hbt!]
	\centering
	\includegraphics[scale=0.4]{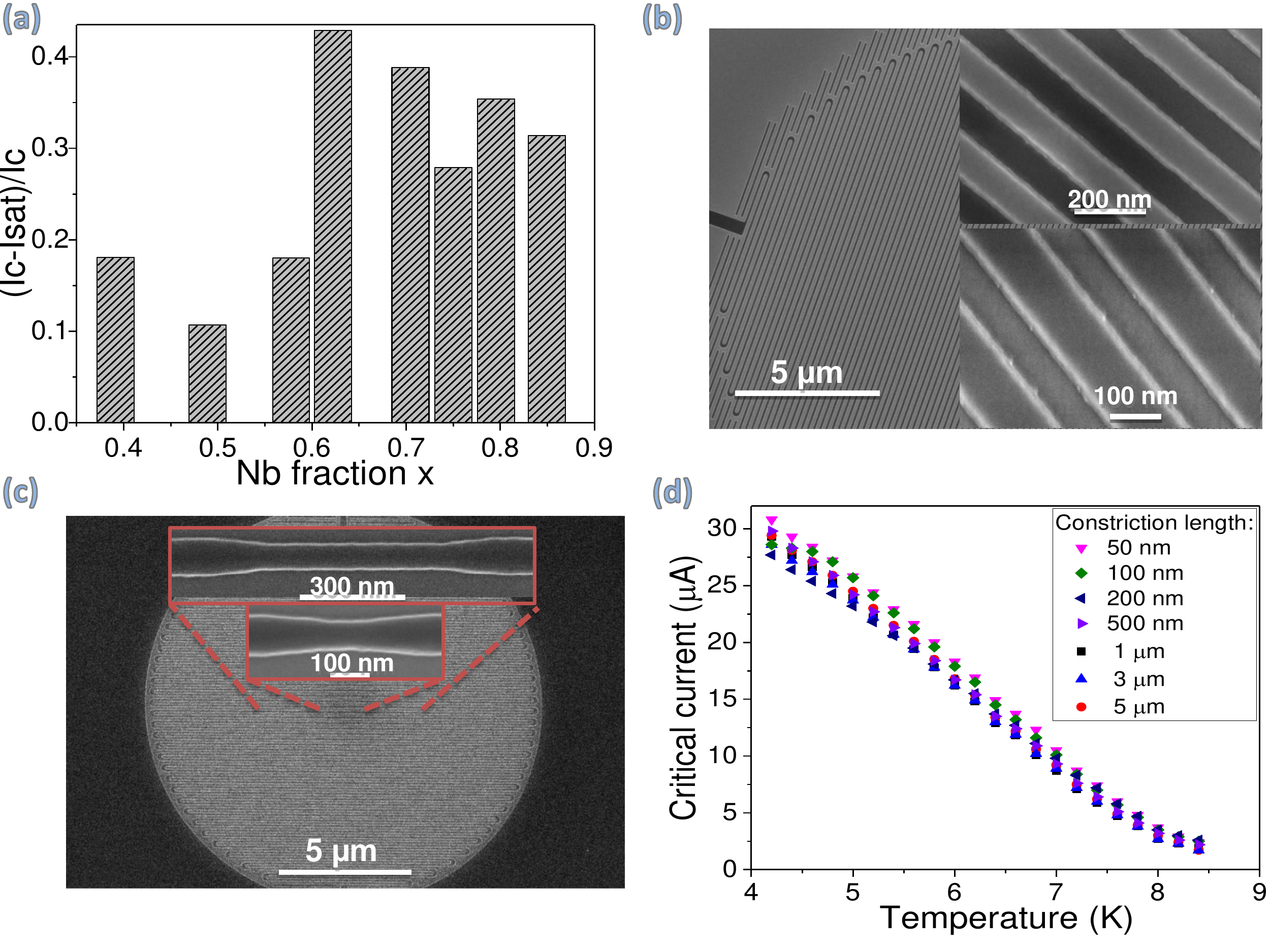}
	\caption{(a) Saturation of internal efficiency for sputtered films of $Nb_{x}Ti(1-x)N$ versus their Nb fraction at 830\,nm. (b) SEM images of devices fabricated and optimized for the wavelength range of 800-900\,nm. Top and bottom insets show devices with filling factor of $\sim$0.55 and 0.6, respectively. (c) Meander with a short constricted region. The main nanowire has a width of 100\,nm and the width in the constriction region is 70\,nm. The constriction length varies from 50\,nm to 5000\,nm and is always a straight section (not including bends). (d) Critical currents of devices described in (c) versus temperature. No significant change of critical currents was observed when changing the constriction length.}
	\label{fig:fig1}
\end{figure}

Magnetron sputtering was used to deposit the films and similar to ref.\cite{Zichi:2019}\,, we varied the sputtering parameters to achieve both high current and saturation of internal efficiency at the wavelength of 820\,nm (we also optimized films and fabricated high performance devices optimized for the telecom wavelength range, see supporting information). Figure\ref{fig:fig1}(a) represents, similar to ref.\cite{Zichi:2019}\,, the measured internal efficiency (the saturation plateau, $\frac{I_c - I_{sat}}{I_c}$) for detectors made from films with different partial Nb contents. As it can be observed, all detectors with different Nb fractions show a saturation plateau, therefore, to enhance the signal to noise ratio and hence the timing jitter, we based our choice on the films which provided the highest critical current density and highest sheet resistance. All devices, unless mentioned, are fabricated from films with a Nb partial ratio of 0.85 ($\mathrm{Nb_{0.85}Ti_{0.15}N}$). This composition yields detectors which not only have a high critical current density and a long saturation plateau (at the studied wavelength), but also can be operated at higher temperatures \cite{Gourgues_highT:2019} and are compatible with large scale integrated nanophotonics \cite{Gourgues_waveguide:2019}. We optimized our fabrication recipe and fabricated detectors with high uniformity. High resolution SEM images of sample devices are shown in Figure\ref{fig:fig1}(b).

Localized inhomogeneities are known to limit the critical current of SNSPDs made out of thin crystalline films (such as NbN and NbTiN) \cite{Gaudio:2014}. We investigate whether the inhomogenities can also affect our thicker NbTiN films by embedding short constrictions in the meandering detector as shown in Figure\ref{fig:fig1}(c). The main SNSPD meander had a width of 100\,nm and the constriction was 70\,nm wide. The length of this constriction was varied from $50\,nm$ to $5\mu m$ (always a straight section wire with no bends) and we measured their critical currents at several temperatures in the range of 4.2\,K to 8.5\,K (Figure\ref{fig:fig1}(d)). We did not observe any significant degradation of critical current with the increase of constriction length. Furthermore, many of our fabricated shorter SNSPDs ($\sim 400 \mu m$ long) and longer SNSPDs ($\sim 4 mm$ long) showed similar critical currents.       

\begin{figure}[hbt!]
	\centering
	\includegraphics[scale=0.5]{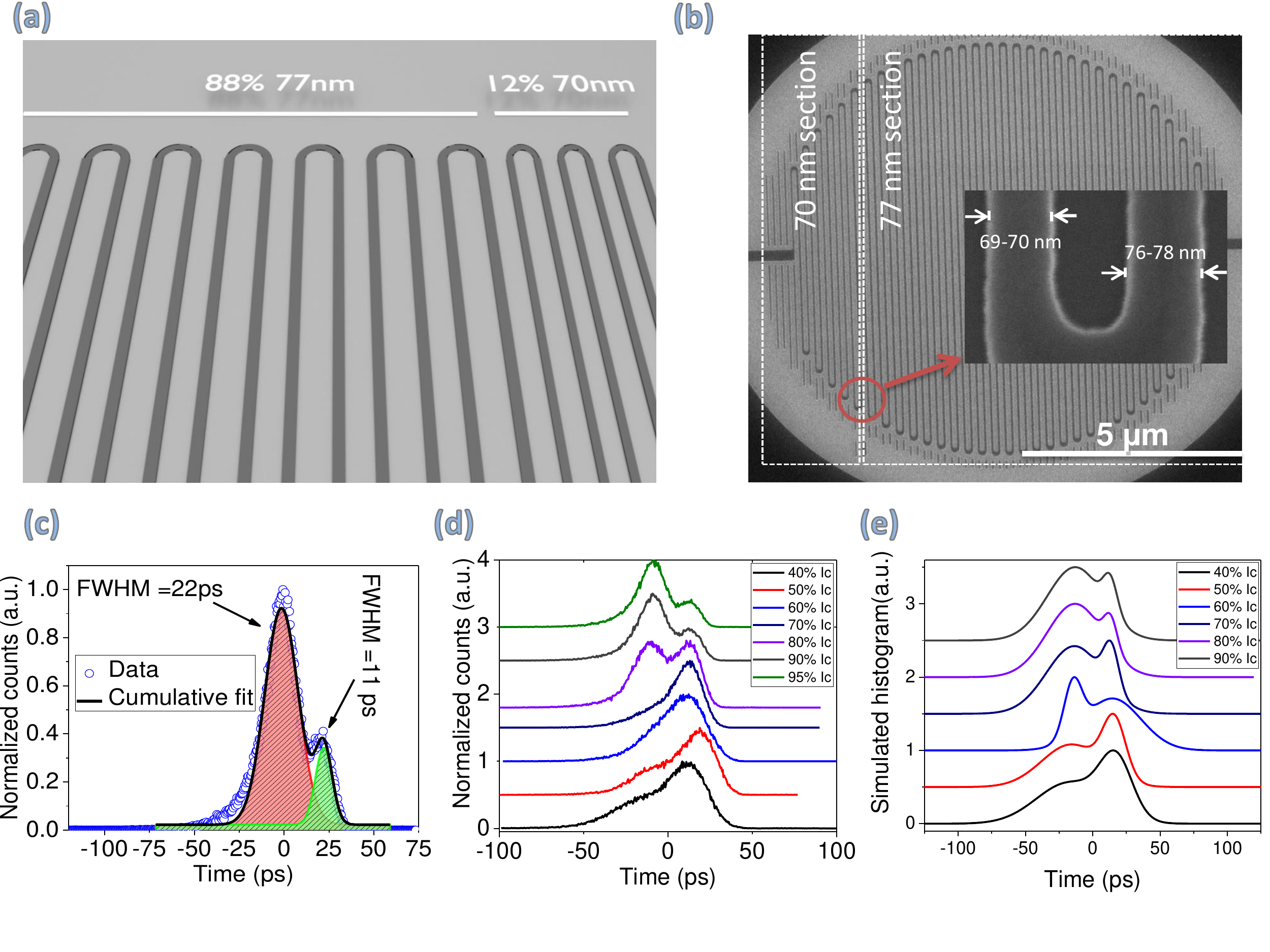}
    \caption{(a) Schematic illustration of a two section SNSPD: having $\sim$12\% (of its area) 70\,nm nanowire width and $\sim$88\% a width of 77\,nm. (b) SEM image of a two section SNSPD. The inset shows a magnified high resolution SEM of the transition between 70 and 77\,nm. The measured values were in good agreement with the design (also see supporting information for more SEM images). (c) Time jitter for the two-section SNSPD, at a bias current equivalent to 95\% of critical current. Two distinct peaks can be observed, which correspond to the two different meander sections. (d) Time jitter measurements for the same detector at various bias currents. It can be observed that at lower biases the detections mostly occur in the 70\,nm section (higher internal saturation) while at higher biases it is dominated by 77\,nm section (larger surface area). This difference depends on the degree of saturation of each section. (e) Simulated correlation histograms at different bias currents.}
	\label{fig:fig2}
\end{figure}

It has been shown that local variation of nanowire width can influence the time resolution of SNSPDs \cite{Connor:2011}. To investigate the relevance of these local variations on the time resolution of our devices, we fabricate and characterize SNSPDs sectioned in two regions with different nanowire widths. Figure\,~\ref{fig:fig2}(a) is an illustration of such a two sectioned SNSPD and Figure\,~\ref{fig:fig2}(b) shows an SEM image of a fabricated device with 70\,nm and 77\,nm wide sections. We designed the device so that the length of 77\,nm wide section is about $\sim$7.7 times longer than the 70\,nm wide section. We measure the SNSPD jitter by correlating the SNSPD pulses (start signal) with that of a pulsed laser (1060\,nm, 5.08\,ps, 48.6\,MHz) as stop signal, shown in Figure\,~\ref{fig:fig2}(c). Two distinct peaks can be observed, corresponding to the two different meander sections. We ascribe the peak on the right side to the 70\,nm wide section because of its higher resistance which leads to faster risetime and hence earlier start signal. The fast risetime also leads to a narrower distribution since noise jitter is inversely proportional to the slope.  The left side peak is ascribed to the 77\,nm wide section which has the lower resistance and hence the late start signal. We also investigated the temporal profile of detection events as a function of bias current, as shown in Figure\,~\ref{fig:fig2}(d). 

To better understand the bias-current-dependent histograms of our two-section device, we numerically simulated these histograms including electronic-noise-induced timing jitter \cite{Zhao:2011,You:2013}, geometric timing jitter \cite{Calandri:2016}, and inhomogeneity-induced timing jitter \cite{Cheng:2017}, see supporting information for details. As presented in Figure\,~\ref{fig:fig2}(e), the simulated histograms show the features of double peaks which were also observed experimentally (Figure\,~\ref{fig:fig2}(d)). The system detection efficiency of each section determines the area under each peak, and therefore, is the major factor affecting the amplitude of each. The separation between the two peaks is the difference of time delays. We see that the simulated separations between the two peaks are larger than the measured results; so that at low bias level, two distinct peaks are obtained in simulation, as opposed to a main peak with a shoulder as observed experimentally. This discrepancy is mainly due to the fact that simulated leading edges of these time-domain pulses do not perfectly match the experimental results (see supporting information). 

Nanofabrication of SNSPDs can introduce imperfections; the nanowire width may vary along the meander for a number of systemic and fabrication issues. A common factor is the proximity effect during electron beam lithography. The proximity effect can cause center areas of the detector to have a wider or narrower width (depending on the tone, thickness and chemistry of the ebeam resist as well as the properties of the substrate, see supporting information for examples of proximity effect simulations), resulting in a similar behaviour to that of Figure\,~\ref{fig:fig2}(c). However, due to the limitations of the experimental setups, it is not always possible to resolve the peaks, yet we observed two distinct peaks in the temporal response of devices with a width difference as low as 2\,nm (supporting information). It should be noted that if the device has more gradual widths variations (which is plausible considering the profile of proximity effects), instead of distinct peaks, one would expect a single broad peak. Proximity effects can be accounted and compensated for, however, it requires simulation, optimization and verification for each single pattern, resist, and every electron beam lithography system. As our simulation suggests (supporting information) and from the literature \cite{PenLi_2015}, that it is possible to significantly reduce the PE using thinner ebeam resists. We tuned the resist thickness (for negative tone, 25-30\,nm XR1541, and for positive tone 90-100\,nm PMMA/AR-P 6200.04), the reactive ion etching chemistry (12.5\,sccm $SF_6$, 3.4\,sccm $O_2$, and a process pressure of $\sim 4 \mu bar$), and etching power and time (50\,W and 45 seconds), and fabricated 70-100\,nm wide nanowires made out of 10-13\,nm thick films. The resulting detectors showed saturated internal efficiency in the wavelength range of 400-1064\,nm. Note that the top section of the film has 0.5-1.5\,nm oxidized layer \cite{Cheng:2016}, preventing it from further oxidation. For high efficiency and high time resolution telecom detectors we used slightly thinner films (8.5-9.5nm). For all fiber-coupled devices, unless otherwise clearly stated, the nanowire width and filling factor were 70\,nm and 0.5, respectively. 

\begin{figure}[hbt!]
	\centering
	\includegraphics[scale=0.45]{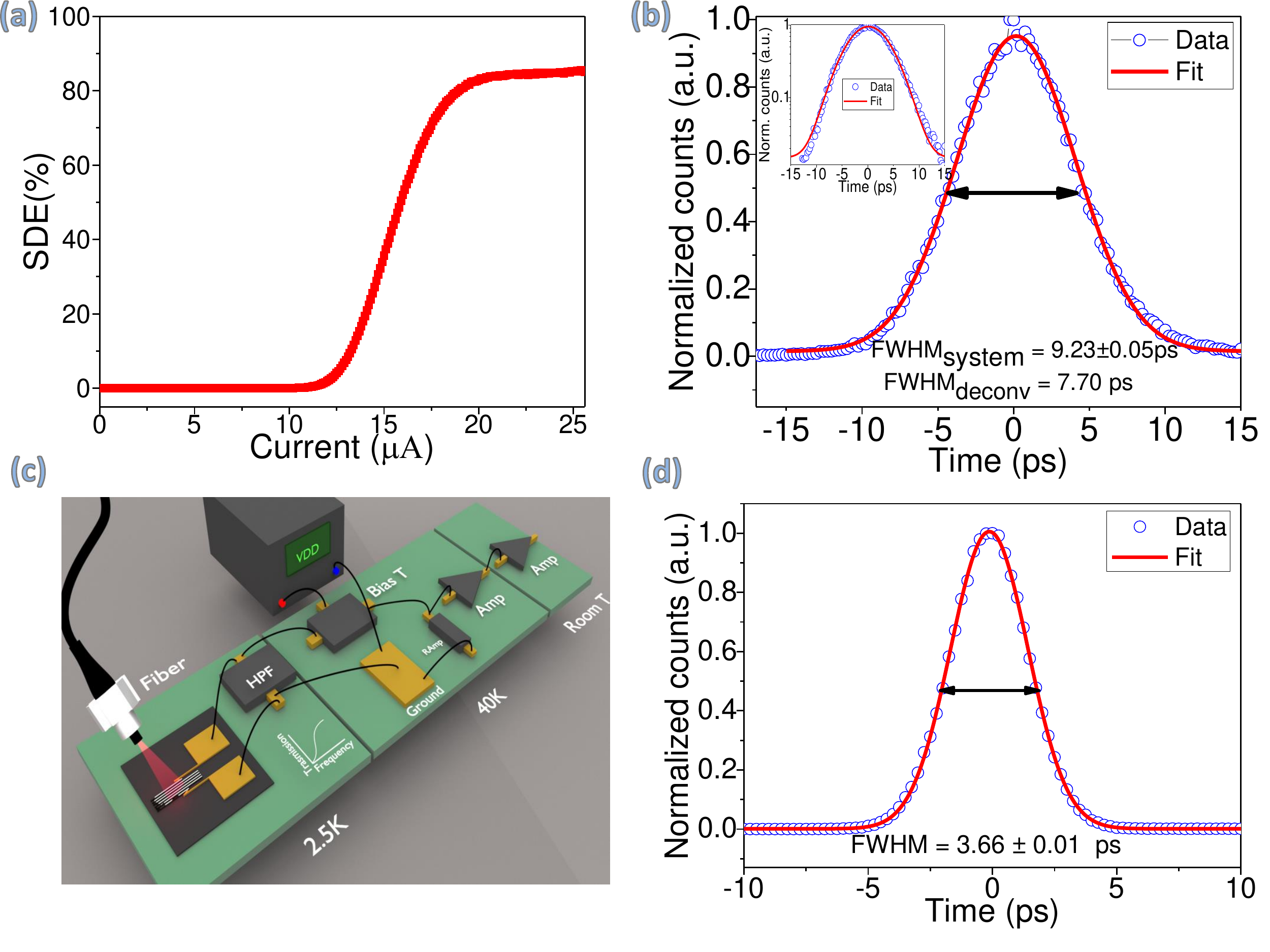}
    \caption{(a) Efficiency vs bias current for a detector at 915\,nm, the peak efficiency is 85$\pm$4.5\% (b) Jitter measurement for the same detector as part (a), the FWHM of the jitter, obtained from a Gaussian fit, is 9.23$\pm$0.05\,ps. Considering the duration of the laser pulses (5.08\,ps) and neglecting the contribution from the correlator and photodiode ($\sim$2.5\,ps), we obtain an SNSPD instrument response function of 7.70\,ps (see supporting information). (c) An illustration of the SNSPD readout circuit used for high photon fluxes. (d) At higher photon fluxes using our new readout circuit, from the fit to the data we calculated a system time resolution of 3.66 $\pm$ 0.01\,ps}
	\label{fig:fig3}
\end{figure}

For the measurement of high performance detectors, we packaged and characterized the detectors in a standard setup similar to \cite{Zadeh:2017}. For efficiency measurements, at each wavelength we adjusted the input power to 10\,nW and then attenuated it by 50\,dB, setting an input photon flux in the range of 200-800\,KCounts/second for visible to telecom wavelength range. For jitter measurements at low photon fluxes, we used a pico-second pulsed laser (same laser as was used for characterization of two-section detectors), and we kept the detector count-rate in the range of 60-100\,KCount/second. Figure\ref{fig:fig3}(a) and \ref{fig:fig3}(b) present the efficiency and jitter measurement results for a detector optimized for the wavelength range of 880-950\,nm, respectively. For other wavelengths we achieved similar performance, see supporting information. We measured a system detection efficiency of $>85$($\pm5\%$) and system time resolution of 9.23$\pm$0.05\,ps (deconvoluted 7.70\,ps, see supporting information). Other devices for this wavelength had SDE of 82-91.5\% and time resolution of 9.5-15\,ps (see supporting information). The inset in Figure\ref{fig:fig3}(b) is a logarithmic plot of the data and fit. On the left side of the jitter measurement data, a small asymmetry and deviation from the fit is observed. Besides originating from the nanowire width variation, this asymmetry could also be due to detection in the nanowire bends as reported in \cite{Sidorova:2017}. However, in comparison with \cite{Sidorova:2017} our degree of asymmetry is smaller, which most likely is explained by the non uniform illumination of our device caused by the fact that the fiber coupled detector has a diameter larger than the mode field diameter of the optical fiber. As a result, the meander bends are not efficiently illuminated. The asymmetry was stronger when we flood illuminated the detectors in a free-space setup (in this case, bends were illuminated). To achieve the best time resolution, we triggered the correlator on the steepest part of the pulses rising edge. We also studied the effect of trigger level (varying it from 10\% to 80\% ) on the value of the FWHM of detector jitter (see supporting information). An increase in the FWHM jitter can be observed, from 15.5\,ps to 22.5\,ps for a sample detector. This is likely to be caused by reduced signal to noise ratio (slower rise time or higher noise level).

We also investigated the count-rate dependence of the jitter (see supporting information). We observed that at low count-rates (below 0.1 $\times$ laser repetition-rate), the SNSPD time jitter exhibits a typical single Gaussian distribution. However, as the count-rate increases, other peaks start to appear in the distribution. The observation of these peaks can be explained by the fact that when the bias recovery time of a SNSPD is longer than $\frac{1}{laser\,rep.\,rate}$, subsequent detection events take place with an under-biased detector while detection events which are temporally further away would experience normal detection (similar to low count-rate regime). Therefore, these different cases result in different risetime/signal levels, which consequently lead to multiple peaks in the jitter measurement.   

Normally, SNSPDs under high excitation powers latch into the normal mode (they become normal with no recovery to the superconducting mode). By providing a low frequency path for the SNSPD current to ground, it is possible to operate detectors at higher powers. Moreover, by using high-bandwidth low-loss coax cables, a low jitter can be achieved. We use a high pass filter, as shown in Figure\,~\ref{fig:fig3}(c), in series with the detector to allow for high power operation together with low jitter. The low-pass filter made it possible to operate the detector with higher powers (1-10\,nW). With higher intensity, we make sure that every single laser pulse is filled with several photons. Therefore, the risetime and signal level stays constant for all detection events and the temporal distance between photons become shorter. Moreover, at these high powers, any SNSPD intrinsic/geometry contribution to the jitter is negligible as only the fastest events trigger the correlator. To this end the only limit we observed was the electrical noise and limited bandwidth of the amplifiers. The result of time jitter measurement with multi-photon excitation (>100 photons per pulse) is shown in Figure\,~\ref{fig:fig3}(d), demonstrating a very high time-resolution of 3.66\,ps ($<$3\,ps after decoupling the contributions from the correlator and the photodiode, see supplementary). 

\section{Quantum Correlation with a Multi-Pixel SNSPD }

Many optical experiments in chemistry, biology and material science deal with scattered light which cannot be efficiently collected and coupled into single mode optical fibers. Therefore, sensors with large active areas are required. SNSPDs with large active areas have gained attention in the recent years and several works on single pixel \cite{Liu:2014,Li:2015,Wollman:2017,Chang:2019} and multi-pixel devices \cite{Allman:2015, Allmaras:2017, Miyajima:2018} have been reported. Compared to Multi-pixel devices, Single-pixel large-area devices are less complex and require only one readout channel. However there is a price to pay, since many performance parameters of SNSPDs such as deadtime and jitter and also fabrication yield suffer as the device length is increased. Until now these factors have limited the work of extending the size of single-pixel SNSPDs. 

\begin{figure}[hbt!]
	\centering
	\includegraphics[scale=0.45]{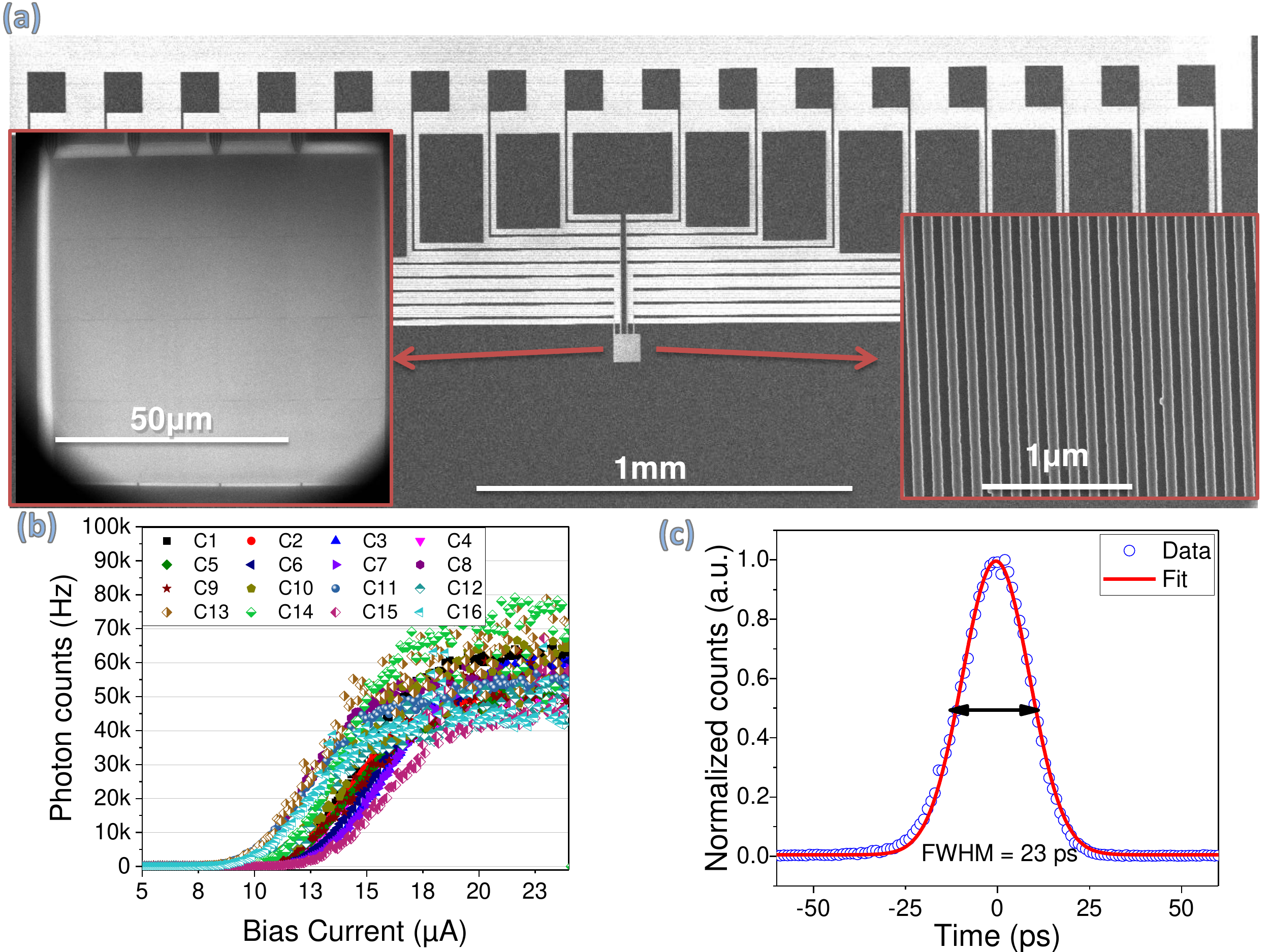}
    \caption{(a) SEM image of a fabricated 16 pixel SNSPD array covering an area of $100\times 100 \mu m^2$ (pixel size $25\times 25$\,$\mu m^2$). The two insets show magnified images of the sensor array. (b) Photon count rate versus bias current for different pixels of the SNSPD array. The performance of different pixels are similar and we attribute the small differences in pixels responses to the fact that the data was collected in two cooldown cycles (temperature and illumination variation). Nevertheless, for a bias current >20$\mu A$, all pixels reach saturation of internal efficiency.  (c) Timing jitter measurement for one of the pixels. The measured system jitter of 23\,ps is >25 times better if we only consider the reported geometrical jitter in other platforms.}
	\label{fig:fig4}
\end{figure}

For multi-pixel SNSPD arrays, as the operation point and performance of individual pixels are usually not uniform, scaling these sensors also pose practical implementation challenges. Here we demonstrate, using our thick NbTiN films, multi-pixel arrays with small variations in performance of the pixels. 
Figure\,~\ref{fig:fig4}(a) show SEM images of a fabricated 16 pixel SNSPD array covering an area of $100\times 100 $\,$\mu m^2$ (pixel size $25\times 25$\,$\mu m^2$). We measure the photon count rate versus bias current for different pixels of the SNSPD array using a flood-illumination setup at 670\,nm. The results are presented in Figure\,~\ref{fig:fig4}(b). It can be observed that the performance of all pixels are similar (for a bias current >20\,$\mu A$, all pixels reach saturation of internal efficiency) and therefore, a much simpler biasing circuit can be used. We attribute the small differences in pixels responses to the fact that the data was collected in two cooldown cycles (cooldown 1: pixels 1-8, cooldown 2: pixels 9-16) and the temperature and illumination might had small variations (also within one cooldown there might be small variations in uniformity of illumination, see supporting information). We also measured the time jitter for some pixels of the array, the results of one such measurement is provided in Figure\,~\ref{fig:fig4}(c). The measured system jitter of 23\,ps for a $\sim$4.46\,mm long detector is >25 times better than other platforms\cite{Kuzmin:2019, Santavicca:2016, Zhao:2018} even considering only the geometrical jitter, which indicates that the electromagnetic field does not have as high confinement in our platform as detectors made from thinner films (capacitive coupling of signal between the lines).

The multi-element detectors are of particular interest in quantum optics as they provide direct photon correlation measurements with no need of spatially splitting the beam to different detectors. We carry out antibunching measurements using a 4-pixel multimode fiber coupled SNSPD on a semiconductor nanowire quantum dot. These sources have been extensively studied in the literature and have demonstrated promising performance in terms of clean emission spectra \cite{Dalacu:2012}, as entangled photon sources \cite{Versteegh:2014} and for near-unity coupling to fibers \cite{Bulgarini_NL_2014}. 
We use a 4-pixel multimode fiber coupled SNSPD to carry photon correlation measurements on the single-photons generated by a nanowire quantum dot. SEM image of the detector is shown in Figure\,~\ref{fig:fig4}(a) with the ground pads (G), signal pads (S), and pixel labels (1, 2, 3, 4) highlighted in the figure. Our measurement setup is illustrated in Figure\,~\ref{fig:fig4}(b), a cryogenically cooled source is excited using a CW laser (516\,nm) and collected photons are filtered using a monochromator (Princeton Instrument Acton 2750) and are coupled to the detector through a graded index $50$\,$\mu m$ fiber. We conduct photon correlation measurements between different pixel pairs of the detector, and the results for 3 of these pairs are presented in Figure\,~\ref{fig:fig4}(c) (unprocessed data are available in the supporting information). The measured low value of $g^2(0)$, consistent or better than previously measured $g^2(0)$ on similar sources \cite{Zadeh_nano_2016}, without using any beamsplitter demonstrates the usefulness and versatility of such detectors.
\begin{figure}[hbt!]
	\centering
	\includegraphics[scale=0.6]{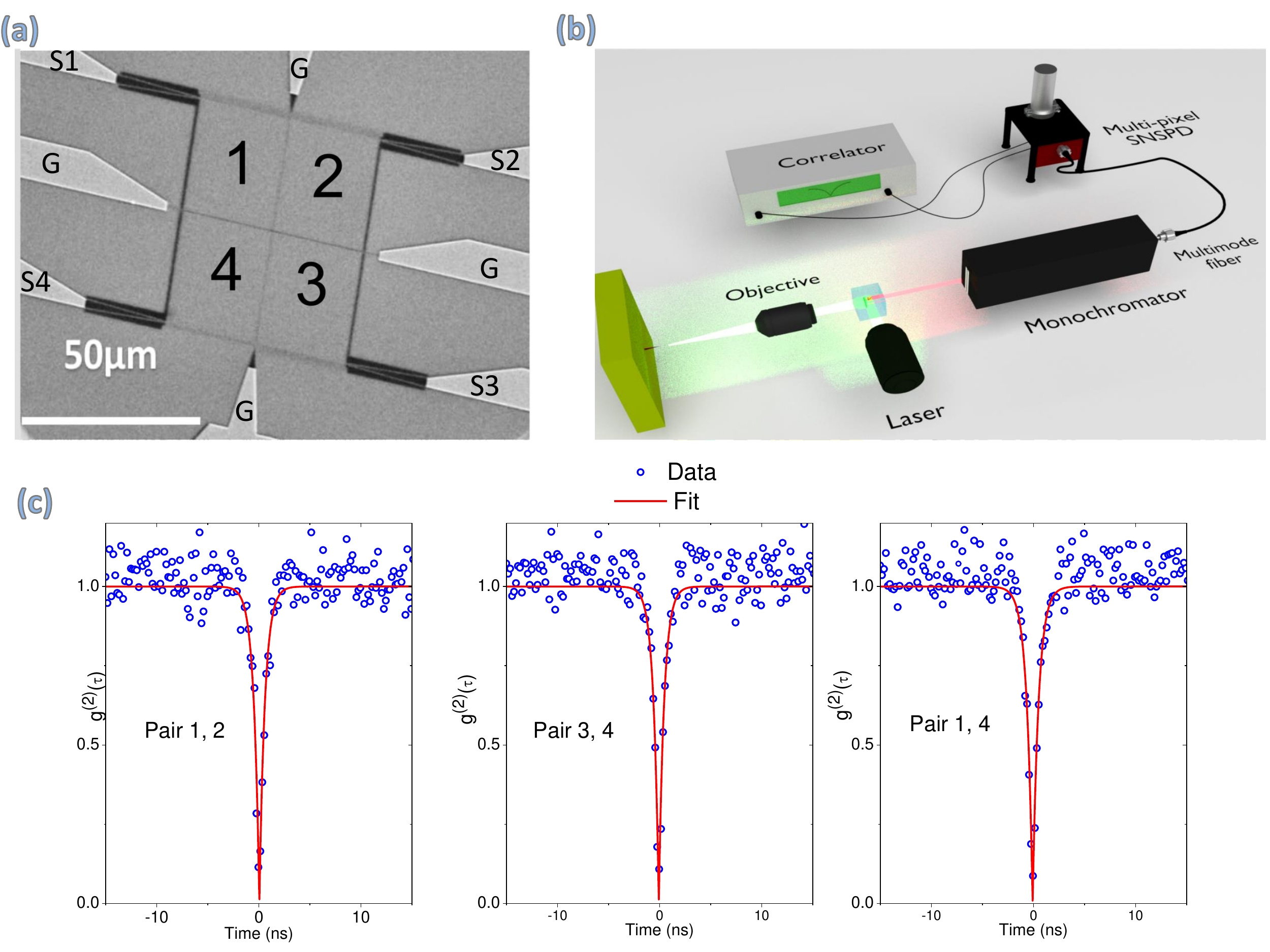}
    \caption{(a) SEM image of a fiber-coupled 4-pixel device covering an area of 50$\times$50$\mu m^2$. (b) Illustration of the photon antibunching measurement setup used to benchmark the 4-pixel detector. (c) Results of photon correlation measurements with three out of six possible pixel pair combinations (complete dataset can be found in supporting information). All pair combinations yield similar outcomes (depending on binning: a $g^{2}(0)$ between 0.03 and 0.1) while the fit at time 0 vanishes completely. The difference between the data and the fit is mainly due to timing jitter induced by a path-length difference in the monochromator (see supporting information).}
	\label{fig:fig5}
\end{figure}

\section{Conclusion}
We demonstrated superconducting nanowire single-photon detectors (SNSPD), fabricated from NbTiN thick films, combining high time resolution and high efficiencies at the same time. The influence of fabrication imperfections on the time resolution of SNSPDs was studied and it was revealed that a small variation in the order of a few nanometer can significantly influence the timing jitter. We showed that thicker films of NbTiN are less susceptible to small local material imperfections and therefore, a good candidate to serve as large active area high performance photonic sensors. A multi-pixel detector was showcased as a quantum correlator in a Hanbury Brown and Twiss experiment without using a beamsplitter. The resulting low $g^2(0)$ demonstrates the potential of such devices not only in the field of quantum optics, but also in the fields such as fluorescence microscopy. Large-area multi-pixel SNSPDs are furthermore expected to have bright prospects in ultra-fast optical and medical imaging applications.

\begin{acknowledgement}
     I. E. Z., A. W. E., V. Z., D. R. S., and Single Quantum B.v. acknowledge the supports from the ATTRACT project funded by the EC under Grant Agreement 777222. I. E. Z. acknowledges the support of Nederlandse Organisatie voor Wetenschappelijk Onderzoek (NWO), LIFT-HTSM (project 680-91-202). R. B. M. G acknowledges support by the European Commission via the Marie-Sklodowska Curie action Phonsi (H2020-MSCA-ITN-642656). A. W. E. acknowledges support from the Swedish Research Council (Vetenskapsr\aa det) Starting Grant (ref: 2016-03905). V. Z. acknowledges funding from the Knut and Alice Wallenberg Foundation Grant "Quantum Sensors", and support from the Swedish Research Council (VR) through the VR Grant for International Recruitment of Leading Researchers (Ref 2013-7152) and Research Environment Grant (Ref 2016-06122). Y. M., K. Z., and X. H. acknowledge the support from Natural Science Foundation of Tianjin City (19JCYBJC16900). 
\end{acknowledgement}

\begin{suppinfo}

The supporting information includes additional examples of high performance devices (S1), supplementary information and simulations for detectors' time jitter (S2) as well as further data on multi-pixel devices(S3).  

\end{suppinfo}

\bibliography{LJHE}
\includepdf[pages={1-14},scale=1]{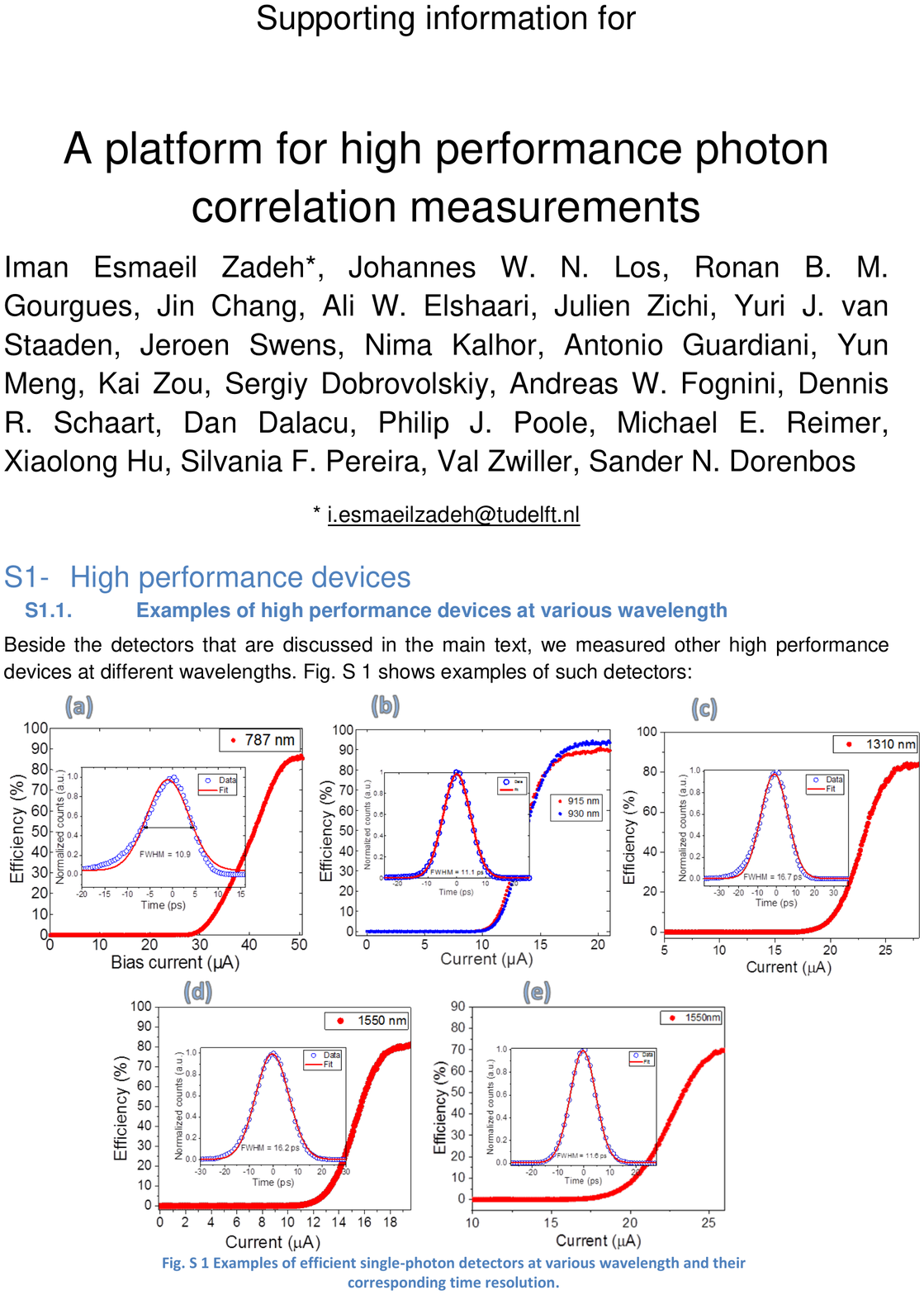}

\end{document}